**Title: Observation of an environmentally insensitive solid state spin defect in diamond**

**Authors:** Brendon C. Rose,[1]* Ding Huang,[1]* Zi-Huai Zhang,[1] Alexei M. Tyryshkin,[1] Sorawis Sangtawesin,[1] Srikanth Srinivasan,[1] Lorne Loudin,[2] Matthew L. Markham,[3] Andrew M. Edmonds,[3] Daniel J. Twitchen,[3] Stephen A. Lyon,[1] Nathalie P. de Leon[1]†

**Affiliation:** [1]Department of Electrical Engineering, Princeton University, Princeton, NJ 08540. [2]Gemological Institute of America, New York, NY 10036. [3]Element Six, Harwell, OX11 0QR, United Kingdom.
* These authors contributed equally to this work
† Corresponding author. E-mail: npdeleon@princeton.edu

**One Sentence Summary:** We report the stabilization and characterization of the neutral silicon vacancy center in diamond, a promising defect for quantum technologies.

**Abstract:**

Engineering coherent systems is a central goal of quantum science. Color centers in diamond are a promising approach, with the potential to combine the coherence of atoms with the scalability of a solid state platform. However, the solid environment can adversely impact coherence. For example, phonon-mediated spin relaxation can induce spin decoherence, and electric field noise can change the optical transition frequency over time. We report a novel color center with insensitivity to both of these sources of environmental decoherence: the neutral charge state of silicon vacancy ($SiV^0$). Through careful material engineering, we achieve over 80% conversion of implanted silicon to $SiV^0$. $SiV^0$ exhibits excellent spin properties, with spin-lattice relaxation times ($T_1$) approaching one minute and coherence times ($T_2$) approaching one second, as well as excellent optical properties, with approximately 90% of its emission into the zero-phonon line and near-transform limited optical linewidths. These combined properties make $SiV^0$ a promising defect for quantum networks.

**Main Text:**

Point defects in diamond known as color centers are a promising physical platform for quantum science and quantum information processing. They are particularly promising candidates for single atom quantum memories to enable quantum networks and long distance quantum communication. As atom-like systems, they can exhibit excellent spin coherence and can be manipulated with light. As solid-state defects, they can be placed together at high densities and incorporated into scalable devices. Diamond is a uniquely excellent host: it has a large bandgap, can be synthesized with sub-ppb impurity concentrations, and can be isotopically purified to eliminate magnetic noise from nuclear spins (*1*). The well-studied negatively charged nitrogen vacancy (NV⁻) center exhibits long electron spin coherence times even at room temperature, and has been used to demonstrate basic building blocks of quantum networks, including spin-photon entanglement (*2*), entanglement with nearby nuclear spins to form quantum registers (*3–5*), remote entanglement of two NV⁻ centers (*6, 7*), quantum teleportation (*8*), and entanglement distillation (*9*). Despite these experimental achievements, the NV⁻ center suffers from a low spin-photon entanglement generation rate that is limited by its optical properties. In particular, only a small fraction of NV⁻ emission is at the zero-phonon line (ZPL), which can be parameterized by its Debye-Waller factor of 0.03 (*10*). Furthermore, NV⁻ exhibits a large static inhomogeneous linewidth (*11*) and significant spectral diffusion (*12*), particularly when placed near surfaces (*13*), which result from a large difference in the permanent electric dipole moment between the ground and excited states. These optical properties severely limit the utility of NV⁻ centers for future scalable technologies.

Recently, the negatively charged silicon vacancy (SiV⁻) center has been demonstrated to have more favorable optical properties. The SiV⁻ center exhibits a large Debye-Waller factor of 0.7 (*14*) narrow inhomogeneous linewidth, and single center linewidths with minimal spectral diffusion (*15, 16*). These narrow linewidths arise from its $D_{3d}$ molecular configuration, as inversion symmetry guarantees a vanishing permanent electric dipole moment, making the optical transition frequency insensitive to electric field noise. These properties have enabled demonstrations of two-photon interference from distinct emitters (*16*) and strong atom-photon interactions in a nanophotonic cavity (*17*). However, orbital relaxation through electron-phonon coupling limits the SiV⁻ electron spin coherence time ($T_2$) to 38 ns, even at cryogenic temperatures ($T = 4.5$ K) (*15*). This results from its imbalanced electronic spin configuration, with total spin $S = ½$ in doubly degenerate orbitals, making SiV⁻ prone to phonon-mediated, dynamic Jahn-Teller-like orbital relaxation (*18*).

These promising demonstrations motivate the search for new color centers in diamond that combine the attractive optical properties of SiV⁻ with the long spin coherence time of NV⁻. A natural approach is to tune the Fermi level of diamond to stabilize a different charge state of the silicon vacancy center in order to access a new spin configuration (*19, 20*) while preserving inversion symmetry. There is some evidence that different charge states of SiV have been observed in uncontrolled samples. Previous work has helped establish a connection between photoluminescence (PL) at 946 nm and the KUL1 defect observed in electron spin resonance (ESR) measurements ($S = 1$, zero field splitting $D = 942$ MHz) (*21–23*) by correlating the PL intensity with ESR transitions across several samples. Charge transfer experiments from SiV⁻ suggest that the KUL1 center is the neutral charge state of the same center, SiV⁰ (*22*).

Here we report the stabilization and characterization of the neutral charge state of silicon vacancy (SiV⁰) by deliberate engineering of the diamond host. We accomplish this by careful control over the concentration of co-occurring defects in the diamond to pin the Fermi level such that the neutral charge state has the lowest formation energy (*24*). Fermi level engineering in diamond is challenging, as it relies on charge compensation between deep defects (*25*), and thus requires careful characterization and control of all defects, including heteroatoms, molecular point defects, lattice damage, and extended defects. Furthermore, the presence of other defects in the lattice can complicate characterization of fundamental properties of color centers, and we show that through careful materials engineering, we are able to uncover highly coherent optical and spin properties of SiV⁰.

To disentangle the contributions of different defects, we first studied a modulation doped diamond (layered sample) that allowed a wide range of relative co-defect concentrations to be accessed in a single sample (Fig. 1B, left). This diamond was grown by microwave plasma-enhanced chemical vapor deposition on a {100} high-pressure high-temperature substrate, and both the boron and silicon concentrations were ramped throughout the growth. The boron precursor was shut off to create a 200 μm low-boron ([B] < 35 ppb) region in the middle of the sample. In this layered sample, we observe emission at 946 nm in bulk PL (Fig. 1A) as well as the KUL1 center in bulk X-band (9.7 GHz) ESR (Fig. 1C). Specifically, we observe four sets of peaks with the external magnetic field aligned along a <111> axis, consistent with the two inequivalent site orientations with splittings that correspond to $S = 1$, $D = 942$ MHz. Furthermore, the hyperfine structure of a single peak is consistent with prior measurements (*21*) (Fig. 1C, inset), and the single set of ¹³C hyperfine peaks is indicative of the inversion symmetry of the center. From the PL and Hahn echo intensities compared to the known Si concentration in the sample, we estimate that only a small fraction of the Si exists as SiV⁰.

The apparent optical and spin properties of SiV⁰ in the layered sample were complicated by heterogeneity and the presence of co-defects. The PL spectrum shows broad emission to the red of 946 nm, as well as three peaks at 952, 975, and 985 nm, associated with other defects (Fig. 1A). Time-resolved pulsed ESR measurements of this sample exhibit multi-exponential decays for both $T_1$ and $T_2$ (Fig. S7), most likely because of dipolar interactions between SiV⁰ centers and interactions with uncontrolled co-defects.

Spatially-resolved PL mapping reveals that 946 nm emission is localized in specific bands in the layered sample (Fig. 1B, bottom left), suggesting that $SiV^0$ may form more efficiently in certain environments. Correlating these regions with spatially-resolved secondary ion mass spectrometry (SIMS) indicates that $SiV^0$ forms in regions with [B] = 1-3 ppm, [Si] = 400 ppb, and [N] below the detection limit of the technique, estimated to be around 1 ppb from prior characterization of growth conditions (Fig. S4). Using this information, we prepared a homogeneous sample with a uniform distribution of $SiV^0$ by implanting $^{28}Si$ into high purity diamond with [B] = 1 ppm, [N] < 5 ppb. A range of implantation energies was used to generate a sample with a large enough signal for bulk ESR and low enough density to avoid dipolar interactions between pairs of $SiV^0$ ($< 10^{16}$ cm$^{-3}$). Before ion implantation, we etch 5-10 μm of diamond to remove subsurface damage resulting from polishing (13). After ion implantation, thermal annealing at 800°C and 1200°C forms $SiV^0$ and repairs implantation-induced lattice damage (13). Both steps are critical for suppressing environmental noise and Fermi level pinning from lattice damage. In contrast to the layered sample, PL mapping of this implanted sample shows a homogeneous distribution of 946 nm emission (Fig. 1B, bottom right), and ESR spin counting indicates a conversion efficiency from implanted $^{28}Si$ to $SiV^0$ of higher than 80% (Fig. S6).

The implanted sample exhibits markedly different behavior in time-resolved ESR compared to the layered sample. Both the electron spin coherence and relaxation times show single exponential behavior, indicating that the $SiV^0$ environment is homogeneous. Below 20 K, the electron spin coherence time measured using a Hahn echo sequence is $T_2 = 0.954 \pm 0.025$ ms (Fig. 2A) and is independent of temperature. This spin coherence time is more than 4 orders of magnitude longer than the $T_2 = 35$ ns reported for $SiV^-$ at 4.5 K (15). This $T_2$ is far below the limit given by the spin relaxation time, which we measure to be $T_1 = 43 \pm 2$ s, which is more than 9 orders of magnitude longer than the orbital $T_1 = 38$ ns reported for $SiV^-$ at 4.5 K (15). The spin relaxation time is also independent of temperature in this range, similar to previous observations of $NV^-$ ensembles (26). The temperature-independent mechanism limiting $T_1$ at low temperatures remains unknown, but the direct (single phonon) relaxation process can be excluded because it would have a $T^{-1}$ dependence.

The decoherence mechanism at low temperature can be inferred by the stretching factor, $n$, extracted from the Hahn echo decay signal $S(t)$ (Fig. 2B) by fitting the data to a stretched exponential, $S(t) = A \cdot \exp\left(-\left(t/T_2\right)^n\right)$. The stretching factor $n = 2$ (Fig. 2A, inset) indicates that the coherence time is dominated by spectral diffusion from the 1.1% natural abundance of $^{13}C$ nuclei in this sample (27), similar to what is observed for $NV^-$ centers (28, 29). Dynamical decoupling with the Carr-Purcell-Meiboom-Gill (CPMG) sequence (30) refocuses decoherence from $^{13}C$ spectral diffusion, extending the coherence time to $T_{2,CPMG} = 255 \pm 20$ ms at 15 K (Fig. 2A, black dots), limited in our experiments by pulse error accumulation.

At temperatures above 20 K, both $T_1$ and $T_2$ rapidly decrease with increasing temperature. The temperature dependence of the spin relaxation is consistent with an Orbach process, $T_1 \propto \exp\left(-E_a/kT\right)$, with an activation energy $E_a = 16.8$ meV. Both $T_2$ and $T_{2,CPMG}$ exhibit a similar temperature dependence to the spin relaxation but scaled by a constant factor of 4100, suggesting that the decay in $T_2$ with temperature is related to the same Orbach process.

We definitively verify the connection between the ESR and PL signatures by scanning a laser across the optical transition while measuring the polarization of the $m_s = 0 \leftrightarrow +1$ spin transition at cryogenic temperatures (4.8 K) (Fig. S9). We observe enhancement and suppression of the spin polarization over a narrow line at 946 nm, as expected for resonant optical pumping. Remarkably, we achieve up to 38% bulk spin polarization using resonant optical excitation (Fig. S9). The high degree of optical spin polarization suggests that there are likely spin conserving and spin pumping optical transitions, which are key ingredients for quantum networks.

We now turn our attention to optical characterization of $SiV^0$. Single $SiV^0$ centers are identified near the surface of the implanted sample using a scanning confocal microscope. A confocal scan with off-resonant excitation (905 nm, 55 mW) and detection in the range of 930-950 nm shows isolated and diffraction-limited PL spots with a peak intensity of ~10 kcounts/s (Fig. 3A). Second-order photon correlation statistics from an isolated spot shows a dip at zero-time delay, $g^{(2)}(0) < 0.5$ (Fig. 3A, inset), confirming that these spots are single photon emitters. Time-resolved measurements of the PL reveal that the emission blinks stochastically (Fig. S11C). This characteristic behavior of single defects indicates that the charge state of $SiV^0$ is unstable under 905 nm illumination, and we observe that the switching can be suppressed with concurrent 532 nm excitation. Taking the photon count rate in the bright state (Fig. S10) and accounting for the low quantum efficiency of the detector (20%), and transmission through the high NA and fiber coupling objectives (53% and 48%, respectively) in this wavelength range, we estimate that this corresponds to a photon emission rate of ~300,000 photons/s. We note that we are unable to fully saturate the $SiV^0$ fluorescence rate with off-resonant excitation, likely a result of the weak phonon sideband and correspondingly weak off-resonant absorption (Fig. S11B). Additionally, we note that bright emitters have a broader inhomogeneous distribution (>20 nm) than the bulk PL linewidth of <1 nm measured in the same sample (Fig. S14). This distribution in wavelength is most likely a result of surface-related strain, and is consistent with recent measurements of $SiV^-$ where inhomogeneous distributions of ~20 nm have been observed in nanodiamonds (*31, 32*).

In contrast to bulk PL in the layered sample, the emission spectra of these single centers exhibit narrow, spectrometer-limited peaks (0.1 nm resolution) with no discernible phonon sideband (Fig. 3C). We estimate a lower bound on the Debye-Waller factor by comparing the intensity at the ZPL with the full bandwidth of the measurement, which includes background and noise. By integrating the measured emission out to 1000 nm and comparing to the integrated intensity at 946 nm, we estimate that 90±10% of the emission is in the ZPL (see supplementary materials). This is in contrast with previous estimates of the Debye-Waller factor of $SiV^0$, which were complicated by the presence of uncontrolled co-defects in this wavelength range (*22*).

We investigate the optical transitions of $SiV^0$ in detail using bulk photoluminescence excitation (PLE) spectroscopy (Fig. 4), in which a narrow linewidth laser (< 200 kHz) is scanned across the $SiV^0$ ZPL while measuring emission in the phonon sideband. Although we are unable to observe a phonon sideband in the PL spectrum using off-resonant excitation, we observe a small but measurable increase in photon counts at wavelengths above 960 nm with resonant excitation at 946 nm. The scan resolves narrow lines with linewidths ranging from 250 to 500 MHz (Fig. 4B), which is a factor of 3-6 wider than the transform limited linewidth of 88 MHz determined by the bulk PL excited state lifetime of 1.8 ns (Fig. 3B). Repeated scans over a 3-hour period show that these lines are completely stable in frequency, showing no measurable sign of spectral diffusion. This is in stark contrast to observations of implanted $NV^-$ centers, which exhibit optical linewidths 10-100 times their natural linewidth (*13*).

This work offers the first identification of a color center in diamond with the combined properties of stable, narrow optical transitions and long spin coherence. In summary, we have developed methods to stabilize the neutral charge state of the silicon vacancy center in diamond, and we have demonstrated that this new center combines favorable spin properties with excellent optical properties, making it a promising candidate for a number of quantum applications. Most importantly, $SiV^0$ exhibits long spin coherence ($T_{2,CPMG}$) approaching 1 s at low temperature, 90% of its emission into its ZPL, near-transform-limited optical linewidth, and no measurable spectral diffusion. These high purity, homogeneous samples will allow for detailed characterization of the fine structure and selection rules of the optical transitions, accurate determination of quantum efficiency, and further characterization of the ground state spin. In particular, on-going investigations include the origin of the Orbach process above 20 K and schemes for dynamic stabilization of the charge state during optical excitation. In addition, $SiV^0$ contains an intrinsic longer-lived quantum memory in the form of the $^{29}Si$ nuclear spin within the defect. Preliminarily, we measure nuclear spin coherence times of $T_{2n} = 0.45 \pm 0.03$ s in a $^{29}Si$ enriched sample (Fig. S8), and future work will explore the limits of this coherence in lower spin density samples.

Furthermore, $SiV^0$ is a particularly attractive candidate for integration with quantum nanophotonic devices. The high conversion efficiency for implanted centers is advantageous for nanoscale patterning and registration (*33*). The emission wavelength at 946 nm is compatible with heterogeneously integrated GaAs photonics (*34*), and efficient, low noise frequency conversion to telecom wavelengths (1550 nm) at the single photon level using three-wave mixing has already been demonstrated at 980 nm (*35*).

Finally, the fabrication of high purity, homogeneous samples enables control over environmental noise and spectroscopy of new quantum systems with highly coherent properties. Our methods for Fermi level engineering using careful control over co-occurring defects can be broadly applied to alter the charge state of other known color centers, such as the germanium vacancy and tin vacancy centers (*36*), as well as for discovery of entirely new color centers in diamond.

**Acknowledgements:** This work was supported by the National Science Foundation under the EFRI ACQUIRE program (grant No. 1640959) and through the Princeton Center for Complex Materials, a

Materials Research Science and Engineering Center (DMR-1420541). D.H. acknowledges support from a National Science Scholarship from Agency for Science, Technology, and Research (A*STAR) of Singapore. The authors gratefully acknowledge Alastair Stacey, Wuyi Wang, Ulrika D'Haenens-Johansson, and Alexandre Zaitsev for advice and help with materials characterization, Nan Yao and John Schreiber at the Princeton Imaging and Analysis Center for help with diamond surface characterization, and Adam Gali, Marc Warner, and Jeff Thompson for fruitful discussions.

**List of Supplementary materials:**
Materials and Methods
Supplementary Text
Figs. S1 to S14
Table S1
Reference 37

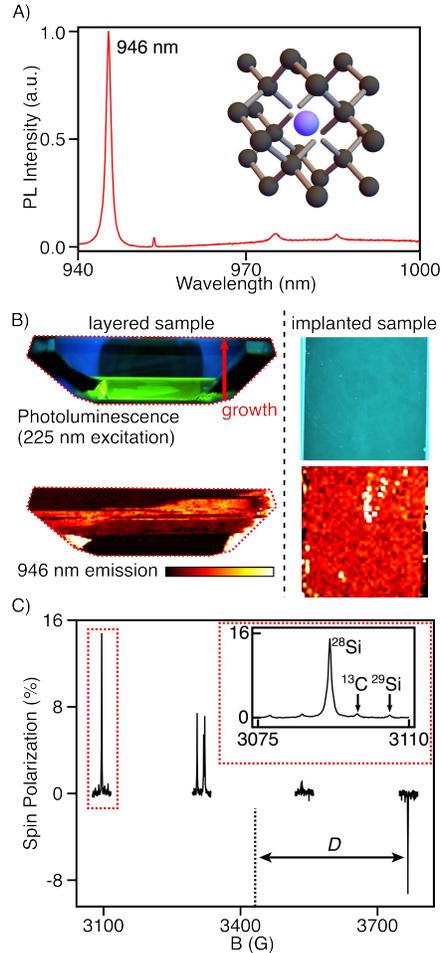

**Fig. 1 Stabilizing SiV⁰ centers in diamond.** (A) Bulk PL spectrum of SiV⁰ at 77 K showing a zero-phonon line at 946 nm. (inset) Ball and stick model of the silicon split-vacancy defect center in diamond. The interstitial Si atom (blue sphere) and split vacancy are aligned along the <111> direction in the diamond lattice. (B) PL with above-bandgap excitation (top) and PL map of emission at 946 nm (colorbar shows intensity in arbitrary units scaled to enhance contrast, yellow indicates high intensity) using 780 nm excitation (bottom) for the two samples under investigation. Layered diamond (left) grown by plasma-enhanced chemical vapor deposition with B and Si concentrations varied during the growth. The growth direction is indicated with a red arrow. PL at 946 nm is localized to a few bands. Implanted sample (right) with uniform boron doping and implanted Si resulting in uniform PL at 946 nm. Saturated intensity (white) corresponds to background and interference arising from the graphitic edges of the sample or broadband PL from the metallic cold finger. (C) Pulsed ESR spectrum at 9.7 GHz with *B* || <111> identifying KUL1 (*S* = 1, *D* = 942 MHz or 336 G) in the uniformly implanted sample. The resonant field for *S* = ½ and a free electron g-factor is indicated as a dashed vertical line in the center of the plot. Four sets of peaks correspond to the two inequivalent sites. Hyperfine structure of an individual transition (inset) shows ¹³C and ²⁹Si peaks. The same pulsed ESR spectrum is observed in both samples.

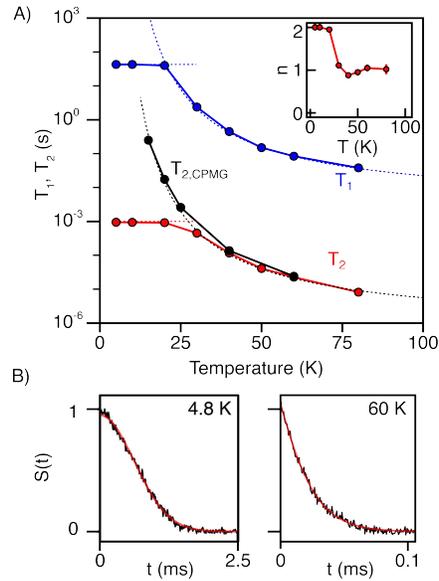

A)

B)

**Fig. 2 Spin relaxation time ($T_1$) and coherence time ($T_2$) measurements**. (A) Temperature dependence of $T_1$ (blue dots), $T_2$ (red dots), and $T_{2,CPMG}$ (black dots) for SiV$^0$ measured on the $m_s = 0 \leftrightarrow +1$ transition of a site aligned with the magnetic field. Below 20 K both $T_1$ and $T_2$ are independent of temperature, with $T_1 = 43 \pm 2$ s and $T_2 = 0.954 \pm 0.025$ ms. Above 20 K both $T_1$ and $T_2$ show a temperature dependence consistent with an Orbach process with an activation energy of 16.8 meV (dashed blue and black lines). Dynamical decoupling with CPMG (black dots) extends the coherence time to $T_{2,CPMG} = 255 \pm 20$ ms at 15 K. (inset) Temperature dependence of the exponential stretching factor, $n$, in the $T_2$ decay showing a step in the stretching factor at 20 K. (B) Selected Hahn echo decay curves illustrating $n = 2$ in the low temperature regime (4.8 K) and $n = 1$ in the high temperature regime (60 K).

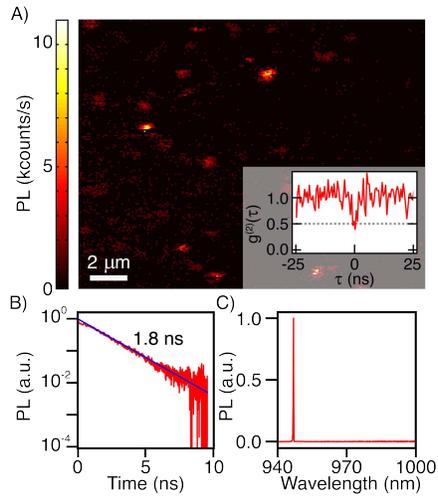

**Fig. 3 Single SiV⁰ centers.** (A) Scanning confocal PL image of isolated near-surface implanted SiV⁰ centers. (inset) Second order correlation function $g^{(2)}(\tau)$ of the PL from a single spot with $g^{(2)}(0) < 0.5$ indicating that the PL is from a single emitter. (B) Time dependent PL using a pulsed laser at 780 nm. The single exponential fit indicates an excited state lifetime of 1.8 ns at 4 K. (C) PL spectrum of a single SiV⁰ center. The emission is collected into a spectrometer limited peak on a CCD spectrometer (0.1 nm resolution), and there is no observable phonon sideband out to 1000 nm.

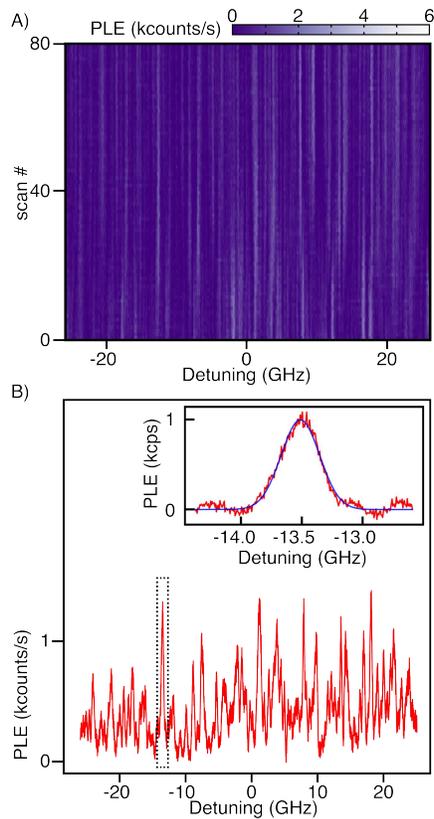

**Fig. 4 Photoluminescence excitation spectroscopy.** (A) A laser is scanned over a 50 GHz range around 946 nm while detecting emission into the sideband (>960 nm). The scan reveals many lines that are stable across 80 iterations spanning a 3-hour period. (B) Emission integrated over time, showing multiple narrow peaks. (inset) Gaussian fit to a single isolated peak, averaged across all 80 scans, with a full width at half maximum of 360 MHz.

# Supplementary Materials for

## Observation of an environmentally insensitive solid state spin defect in diamond

Brendon C. Rose, Ding Huang, Zi-Huai Zhang, Alexei M. Tyryshkin, Sorawis Sangtawesin, Srikanth Srinivasan, Lorne Loudin, Andrew M. Edmonds, Matthew L. Markham, Daniel J. Twitchen, Stephen A. Lyon, Nathalie P. de Leon*

*correspondence to: npdeleon@princeton.edu

**This PDF file includes:**

**Materials and Methods**

<u>1.1  Confocal microscopy measurement setup</u>

Optical measurements were carried out on a home-built scanning confocal microscope optimized for near infrared (nIR) excitation and detection (Fig. S1) with a Janis ST-500 helium flow cryostat and a 0.80 NA, 100x objective (Zeiss EC Epiplan 100x). The focal plane of the objective was controlled with a piezoelectric stack (ThorLabs PK3JRP2), while the lateral confocal scan was controlled with dual-axis scanning galvo mirrors (ThorLabs GVS012). The microscope contains three excitation channels and one detection channel, depicted with different colors in Fig. S1. The confocal scans (Fig. 3A) were performed with off-resonant excitation using a 905 nm diode laser (Qphotonics QFLD-905-200S) in the red channel. Time dependent photoluminescence measurements (Fig 3B) utilized a 780 nm pulsed laser (Toptica FemtoFErb 780) exciting in the red channel. Far off-resonant excitation with a 532 nm diode laser (Coherent Sapphire LP 300) in the green channel was used to stabilize the photoluminescence switching of the near surface SiV$^0$ centers (Fig. S11). The red and green channel were combined with a dichroic beam splitter (ThorLabs DMLP650). Resonant optical excitation was carried out by scanning an external cavity diode laser (New Focus Velocity TLB-6700 laser) between 940 nm and 950 nm (blue channel).

The detection path (yellow) was combined with the resonant excitation path through a 50:50 beam splitter. The yellow and blue paths were coupled to the off-resonant excitation channels via a second dichroic beam splitter (Semrock FF925-Di01). The detection path was filtered from leaked excitation and Raman lines (Semrock FF01-937/LP25 and Thorlabs FESH1000). Fluorescence from the SiV$^0$ phonon sideband was collected with flip-mounted longpass filters (Semrock LP02-980RE-25 and Semrock LP02-980RU-25). The filtered fluorescence signal was coupled into a 50:50 fiber beam splitter, and split to a fiber-coupled single photon counting avalanche photodiode (Excelitas SPCM-AQRH-44-FC) as well as a CCD spectrometer (Princeton Instruments Acton SP-2300i with Pixis 100 CCD and 1200 g/mm grating) with a resolution of 0.1 nm. The photon autocorrelation measurement (Fig. 3A, inset) was accomplished by replacing the spectrometer with a second single photon counting module detector (Excelitas SPCM-AQRH-12-FC) and measuring arrival times using a time-correlated single photon counting module (PicoQuant PicoHarp 300).

<u>1.2  Detector calibrations in the nIR</u>

We characterized the detection efficiency of the single photon counting avalanche photodiode against a photodetector with known responsivity (Newport Model 918D-IR-OD3R with a NIST-traceable calibration). The detection efficiency calibration data are taken with a narrow linewidth external cavity diode laser (New Focus Velocity TLB-6700 laser) that is fiber-coupled into the photodiodes. We first determined the transmitted laser power out of the fiber as a function of wavelength by measuring the current output of the calibrated detector. The same scan was then repeated under the same setup conditions for the avalanche photodiode. We used the known power into the detector from the calibration scan to get an incident photon rate into the detector and to compute the detection efficiency of the avalanche photodiode in our detection wavelength range (Fig. S2A).

We reported the photoluminescence spectrum for a single $SiV^0$ at 4.2 K (Fig. 3C). The raw data was adjusted using a calibrated spectrum taken with a white light source (quartz tungsten halogen lamp) in this range (Fig S2B). To obtain the calibrated spectrum, the white light source was placed at the position of the sample to account for the wavelength dependent loss in the detection channel. From the corrected spectrum, we calculate a Debye-Waller factor of 90±10%.

1.3  Photoluminescence mapping and DiamondView measurement setups

The photoluminescence maps in Fig. 1B (bottom) were taken using a Raman microscope (ThermoFisher DXRxi) modified to include several excitation sources and a back illuminated CCD detector (ANDOR Newtown EMCCD DU970P). A 200 mW 780 nm laser was used for excitation. The samples were cooled to 77 K using either a liquid nitrogen bath or a liquid nitrogen flow cryostat (Linkam THMSG600).

The diamond images in Fig. 1B (top) under above-bandgap illumination were taken at room temperature in a diamond grading tool (IIDGR DiamondView). In this tool shortwave UV light is filtered to wavelengths shorter than 225 nm and supplied to the diamond. This provides above-bandgap light which causes many defects to fluoresce or phosphoresce, and the emission is imaged onto a CCD camera.

2.  Electron spin resonance measurement setup

Electron spin resonance was performed on a modified Bruker Elexsys 580 system with a 1.4 T electromagnet (Fig. S3). This is done by driving the TE011 mode of a cylindrical dielectric resonator (ER-4118X-MD5) (9.7 GHz) with an external vector microwave source (Agilent E8267D). Temperature dependent measurements were performed in an Oxford CF935 helium-flow cryostat. The microwave excitation channel consists of a vector microwave source (Agilent, E8267D), a variable gain solid state amplifier (AR 20S4G11), a protection PIN diode (Hittite switch), and a variable attenuator (ARRA P4844-30). This allows for arbitrary waveform control as well as flexibility in the microwave power delivered to the resonator which is necessary for handling the range of quality factors and excitation bandwidths needed across several samples. The microwave excitation is coupled capacitively to the resonator through a microwave waveguide terminated with an open-loop antenna and the reflected microwaves are returned through a fast Hittite switch (HMC547) redirecting the microwaves to the detection channel. The reflected microwave signal is sent through a cryogenic low noise amplifier (Low Noise Factory, LNF-LNC4_16B) and then through a room temperature low noise amplifier (Bruker). To avoid saturating and damaging the cryogenic amplifier it must be protected from the microwave excitation during the ring down period of the resonator. This protection was supplied by the fast Hittite microwave switch, which was used to couple the reflected microwave excitation out of the resonator back to the excitation pathway.

After the ring down period the switch is opened up to the cryogenic amplifier so that any spin signal outside of the ring down period is amplified. The amplified signal is sent to a quadrature detector (Bruker) where it is mixed with a reference signal from the vector source. The in-phase (I) and quadrature (Q) signals coming out of the mixer are read with a fast digitizing board (Bruker SpecJet). Triggering in this experiment was controlled with a PulseBlaster ESR-PRO board.

**Supplementary Text**

## 1. SIMS Analysis of the layered sample

Combining PL mapping and SIMS allowed for the determination of the locations at which SiV$^0$ was most readily formed, correlated with the elemental compositions in those regions. In the SIMS analysis samples were taken in 8 second intervals and averaged by taking 10 samples at each position. After this averaging, the noise floor of the measurement was in the range of $5x10^{15}$ cm$^{-3}$. This averaging was done at several points in 50 µm steps along a 1050 µm path length along the growth direction.

The initial points of the profile (from 0 to 300 µm) were taken in the HPHT region of the diamond (green region in Fig. 1B) which showed a low concentration of both silicon and boron, in the noise floor of the measurement. The CVD growth starts at 300 µm and continues until 1050 µm. During the CVD growth, both the boron and silicon precursors were ramped in concentration, this resulted in an exponential increase in the elemental concentrations of both boron and silicon in the diamond. The silicon concentration varies across the profile from $2x10^{16}$ cm$^{-3}$ (0.1 ppm) to $2x10^{17}$ cm$^{-3}$ (1 ppm). The boron concentration varies from $1x10^{17}$ cm$^{-3}$ (0.5 ppm) to $4x10^{17}$ cm$^{-3}$ (2 ppm), except for a 200 µm region in the middle of the growth where the boron precursor was suddenly switched off resulting in a boron concentration of $1x10^{16}$ cm$^{-3}$ (0.1 ppm).

## 2.1 Implanted sample recipe

Implantation and annealing was used to generate a homogeneous sample. Before ion implantation, boron was introduced into the diamond during the growth process with a concentration of $10^{17}$ cm$^{-3}$. We etched 5-10 µm of diamond to remove subsurface damage resulting from polishing. The surface of the diamond was strain-relief etched using an inductively coupled plasma reactive ion etcher (ICP-RIE, PlasmaTherm), first with Ar/Cl$_2$ chemistry followed by O$_2$ (12). In order to optimize the bulk spin signal while avoiding dipolar interactions between spins, the implantation volume (depth) was maximized by multi-step implantation. The maximum implantation energy commercially available was 400 keV (Innovion), corresponding to a Si implantation depth of ~450 nm. A total of seven implantation steps (Table S1) were used to generate a uniform distribution up to this maximum depth with a total fluence of [$^{28}$Si] $=3.0x10^{11}$ cm$^{-2}$ and a volume density of ($7x10^{15}$ cm$^{-3}$). Vacancies from knock-on damage due to the implanted high energy $^{28}$Si were simulated using the Stopping Range of Ions in Matter Monte Carlo simulation package (SRIM) (24) and estimated to be ~$1x10^{16}$ cm$^{-3}$. After implantation, high temperature vacuum annealing ($<10^{-6}$ Torr) was performed in three steps: 400 $^o$C for 8 hours to move interstitial defects, 800 $^o$C for 8 hours to form SiV, and 1200 $^o$C for 2 hours to eliminate divacancies and multivacancies. Finally, refluxing in 1:1:1 concentrated sulfuric, perchloric, and nitric acids removed graphitic carbon formed during thermal annealing.

## 2.2 Determining the activation yield of implanted SiV$^0$ with spin counting

The number of SiV$^0$ centers formed in the implanted sample was determined in a spin counting experiment (Fig. S6) using pulsed ESR. In this experiment the amplitude of the SiV$^0$ Hahn echo signal in the implanted sample was compared to the amplitude of the Hahn echo signal from a known reference sample. The reference sample was an isotopically enriched $^{28}$Si crystal doped with phosphorus. The number of phosphorus donor electron spins in this sample was known accurately from measurements of the instantaneous diffusion time (37). The two samples were measured under the same setup conditions including temperature and resonator Q-factor values. The interpulse delay, $\tau$, in the Hahn echo experiment was set much shorter than $T_2$ in both samples. To make a direct comparison, no optical spin polarization was induced in this measurement. Instead, the spins were allowed to reach Boltzmann equilibrium by using a long delay between repeated experiments (repetition time much longer than $T_1$). The echo strength then corresponds to the Boltzmann equilibrium spin polarization (4.8 %) at 4.8 K and 9.7 GHz. Considering the difference between the number of sites, transitions, and hyperfine lines as well as the difference in the dipole strength between a spin-1 species and a spin-1/2 species we evaluated the total number of spins in the implanted sample. This measurement was performed in two separate runs, and the resulting conversion efficiency was estimated to be 90±10%.

## 3.1 Electron spin resonance measurements of the layered sample

The spin coherence times, spin relaxation times, and spin polarization for the layered sample were measured at cryogenic temperature (Fig. S7A). We did not observe single exponential decays in relaxation or coherence, indicative of the wide distribution of SiV$^0$ concentrations present in this sample. As an estimate of the time scales, we fit each decay to a biexponential, $S(t) = A_f \exp(-t/T_f) + A_s \exp(-t/T_s)$, with fast and slow parameters $A_f$, $T_f$ and $A_s$, $T_s$, respectively (Fig. S7B). The shaded regions in panel A represents the interval $[T_f, T_s]$ as a function of temperature. The red region corresponds to the distribution in the spin coherence $[T_{2f}, T_{2s}]$ and the blue region corresponds to the distribution in the spin relaxation $[T_{1f}, T_{1s}]$. These regions are plotted along with the $T_1$ and $T_2$ curves for the uniformly doped sample (from Fig. 2 in the main text). We can see that a similar decrease in $T_1$ and $T_2$ occurs in both samples around 30 K. The saturated $T_{2s}$ = 410 µs at lower temperature is shorter in the layered sample compared to the implanted sample, likely because of instantaneous diffusion from the locally large concentration of SiV$^0$ in that sample observed by PL mapping. The decoherence and relaxation rates are enhanced in this sample when cooling down from 5 K to 1.5 K, in contrast with the implanted sample. The saturated optical spin polarization induced with 532 nm light is plotted as a function of temperature in Fig. S7C for both the implanted sample (red curve) and the layered sample (black curve). The spin polarization was measured for an SiV$^0$ site with its symmetry axis (<111>) oriented along the magnetic field. We achieve a maximum of 13% spin polarization at 5 K in the implanted sample which is consistent with previous observations (21).

## 3.2  Electron nuclear double resonance measurements of nuclear spin coherence time.

$SiV^0$ also contains an intrinsic longer-lived quantum memory in the form of the $^{29}Si$ nuclear spin within the defect. We measure the nuclear spin coherence (Fig. S8, $T_{2n}$) of the intrinsic $^{29}Si$ nuclear spin ($I = 1/2$) in a separate sample that was isotopically enriched during CVD growth with 90% $^{29}Si$ and a $SiV^0$ concentration of 340 ppb. Using electron-nuclear double resonance at the $m_s = 0 \leftrightarrow +1$ transition we observe $T_{2n} = 0.45 \pm 0.03$ s, with a flat temperature dependence at low temperature ($T < 15$ K). The nuclear spin coherence is limited by direct flip-flops between pairs of $SiV^0$ electron spins at this concentration (37), indicating that the nuclear spin coherence can be extended by decreasing the concentration of $SiV^0$.

## 4.  Optical spin polarization with resonant optical excitation

The connection between the zero-phonon photoluminescence line at 946 nm and the KUL1 ESR center was verified by scanning a narrow linewidth laser around 946 nm while detecting the polarization induced on one of the $SiV^0$ electron spin transitions. The optical spectrum of the induced spin polarization was measured by orienting the crystal with B ∥ <111> and measuring the strength of the integrated Hahn echo signal from the $m_s = 0 \leftrightarrow +1$ transition of a site parallel to the magnetic field (Fig. S9A, top) as well as for a site oriented at 109° relative to the magnetic field (Fig. S9A, bottom). The tunable laser was shuttered during the microwave sequence to avoid illumination-induced dephasing. The polarization generated by the laser took almost 30 seconds to saturate with 40 mW incident on the sample (Fig. S9B). The maximum polarization obtained was 38%, accomplished by exciting at 947 nm. The saturation curve for the 947 nm excitation shows a biexponential behavior; this is possibly due to the presence of two overlapping transitions at this excitation wavelength. A small amount of spin polarization can also be observed by exciting at 952 nm, where the PL spectrum (Fig. 1A) shows a small peak. However, the polarization is about an order of magnitude smaller. The nature of this process is still unclear.

## 5.1  Off resonant optical illumination of isolated emitters

The isolated bright surface emitters that appear in a confocal scan under the presence of 905 nm excitation (Fig. 3A) exhibit a wide distribution in the frequency of their zero-phonon line (Fig. S10A). This distribution is much wider than the 0.5 nm linewidth of the ZPL measured in bulk photoluminescence for the $^{29}Si$ enriched sample at 4 K (Fig. S12) or the 1 nm bulk linewidth of the implanted sample at 77 K (Fig. S14). The distribution of bright emitters represents a biased sub-ensemble of the bulk distribution that was preselected by their brightness in the confocal scan. It is likely that this brightness is correlated with high strain. Similarly wide distributions of bright-strained SiV- centers have been observed in nanodiamonds (30, 31).

We were unable to fully saturate the count rate of the emitters. The saturation curve shown in Fig. S11B plots the count rate of an individual SiV$^0$ (taken from the histogram in Fig. S10, left) near the surface of the diamond as a function of the incident laser power (transmitted through the objective) with 905 nm excitation. The red line in this plot is a least-squares fit of the saturation function $R=R_{sat} P/(P+P_{sat})$, where $R$ is the photon count rate at a given excitation power $P$, $R_{sat}$ is the saturated count rate as $P\rightarrow\infty$, and $P_{sat}$ is the saturated excitation power. The fit gives $R_{sat}$=13.2 kcounts/s and $P_{sat}$=53 mW with a goodness of fit parameter $r^2$=0.932.

In addition to the single center described in the main text, we took saturation histograms of a second center that was significantly brighter (S10, left). We can make a conservative estimate of the actual saturated count rate by accounting for the transmission losses (no coupling losses included) in our setup at 946 nm as well as the quantum efficiency of our detector. These losses are dominated by the transmission through the microscope objective (53%), beamsplitters (45%, 55%), and fiber coupler (48%) as well as the measured quantum efficiency of the silicon APD (22%). Combining these factors gives an estimated ~10$^6$ photons/s for the saturated photon emission rate of SiV$^0$ collected by the objective, which is consistent with the short excited state lifetime for this center.

The isolated single center fluorescence we observe near the surface of the diamond in confocal photoluminescence measurements exhibits binary switching in the count rate (Figs. S10 and S11C-D). For a single emitter under illumination of ~50 mW of 905 nm laser light, the average switching time was 300 ms between an "on" state with ~2 kcounts/s and an "off" state with ~600 counts/s (dark count rate). The distribution in the single center photoluminescence rate, shown in red in Figs. S10 and S11D, has a bimodal character. The switching characteristics depend on the power (Fig. S10) as well as the wavelength of the excitation source. Adding an additional 5 mW of green excitation to the 905 nm excitation collapses this distribution into a single peak (green time trace and green distribution) with an average count rate that lies between the two peaks of the bimodal distribution. It is plausible, however, that the switching is still present but is much faster and unresolved in this measurement, which would be consistent with the fact that the maximum count rate under the additional green illumination is smaller than the maximum count rate with only 905 nm excitation.

## 5.2  Resonant optical illumination

In the photoluminescence excitation experiment (PLE) a narrow linewidth laser (Fig. S1) was tuned across the optical transition of the SiV$^0$ center while reading out the counts in the phonon sideband. We measure count rates in the sideband region up to 6 kcounts/s (Fig. 4A) coming from a single center. The transmission losses in this measurement in this wavelength region are mainly due to the objective (53 %), fiber coupler (48 %), 50:50 beamsplitter (55 %), and the silicon APD (~15%). Combining these factors give an estimated ~250x10$^3$ photons/s collected by the objective.

## 6.1  Sample-to-sample variation of phonon sideband measurements

In uncontrolled samples, a direct measurement of the Debye-Waller factor is obscured by the presence of co-defects that emit into the same spectral region as the phonon sideband. This can result in an underestimate of the Debye-Waller factor and variation in the inferred value across samples. Such co-defects have been previously reported near $SiV^0$ (21): emission in the sideband region is observed to increase as the sample is annealed at high temperatures. We observe such sample-to-sample variation when comparing the bulk PL spectra of the layered sample (Fig. 1A) with the $^{29}Si$ sample (Fig. S12). The apparent Debye-Waller factor of the $^{29}Si$ sample is only 9%, an order of magnitude smaller than the factor for the layered sample, most likely because of the presence of other uncharacterized defects in this emission band.

For the single center PL spectrum (Fig. 3C) we do not observe any phonon sideband, limited by the noise floor of the measurement. We know from PLE spectroscopy (Fig. 4) that there must be some photoluminescence above 960 nm. We make a conservative estimate of this fraction based on the signal to noise of the measurement by assuming that there is an unresolved sideband signal equal to the root mean square noise level and spanning a range of 50 nm, comparable to the width of the phonon sideband of $SiV^-$. The result gives an estimate that 90% ± 10% of the counts are in the ZPL.

6.2  Temperature dependence of $SiV^0$ photoluminescence and excited state lifetime

The temperature dependence we observe for the 946 nm $SiV^0$ ZPL photoluminescence (Fig. S13) is consistent with phonon broadening as temperature is increased. This broadening effect saturates at lower temperature, which is inconsistent with previous reports which showed the overall intensity decreasing with decreasing temperature in the range 5 K to 100 K (21). This discrepancy is likely a result of composition variation among samples, and highlights the advantages of engineering samples with the desired color center rather than relying on unique samples with unknown composition. The excited state lifetime (Fig. S13C) is independent of temperature within the range of our measurement, 5 K to 40 K.

6.3  Bulk photoluminescence of implanted sample

Bulk PL of the implanted sample at 77 K (Fig. S14). The $SiV^0$ peak has a 1.0 nm linewidth which is consistent with the linewidth of the $^{29}Si$ sample at the same temperature (Fig. S13). The large background arises from the Raman peak associated with 830 nm excitation.

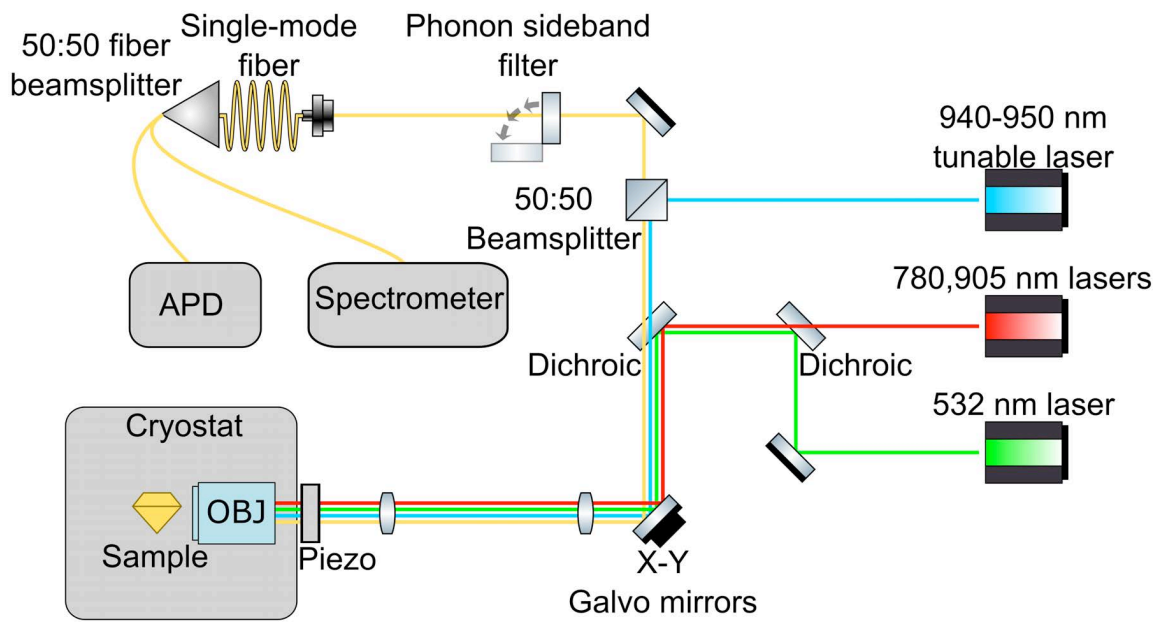

**Fig. S1 Confocal microscopy experimental setup**

Diagram of the nIR scanning confocal microscope. The three excitation channels (green, red, and blue) couple light ranging from visible wavelengths to nIR which enables the far off-resonant PL as well as resonant PLE to be performed in the same setup. The sample sits inside of a variable helium flow cryostat with a high NA objective. The detection path can be coupled to either a silicon APD or a silicon CCD spectrometer.

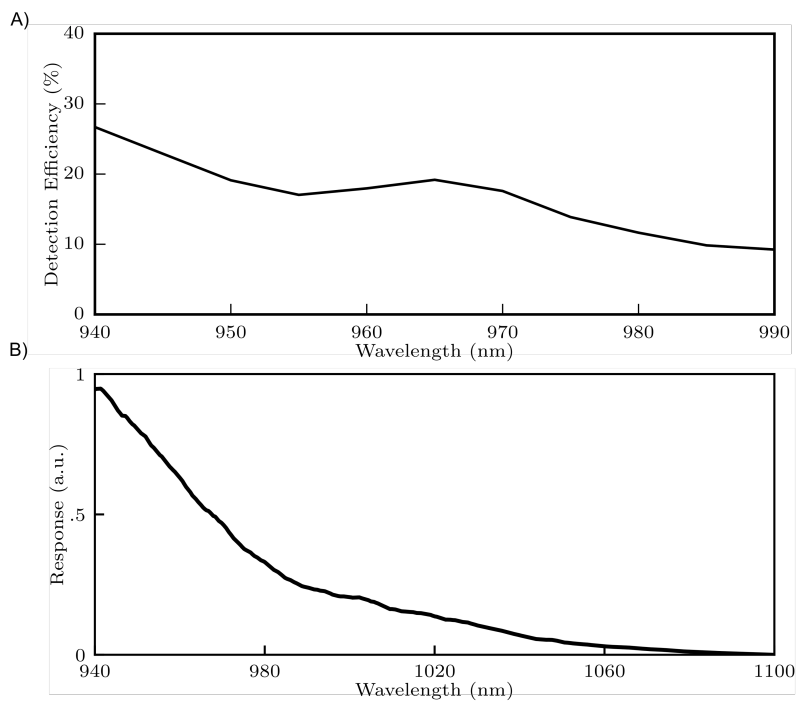

**Fig. S2 Detector calibrations in the nIR**

(A) Photon detection efficiency of the silicon APD detector in the range 940-990 nm. (B) Wavelength response function of the silicon CCD spectrometer in the range 940-1100 nm.

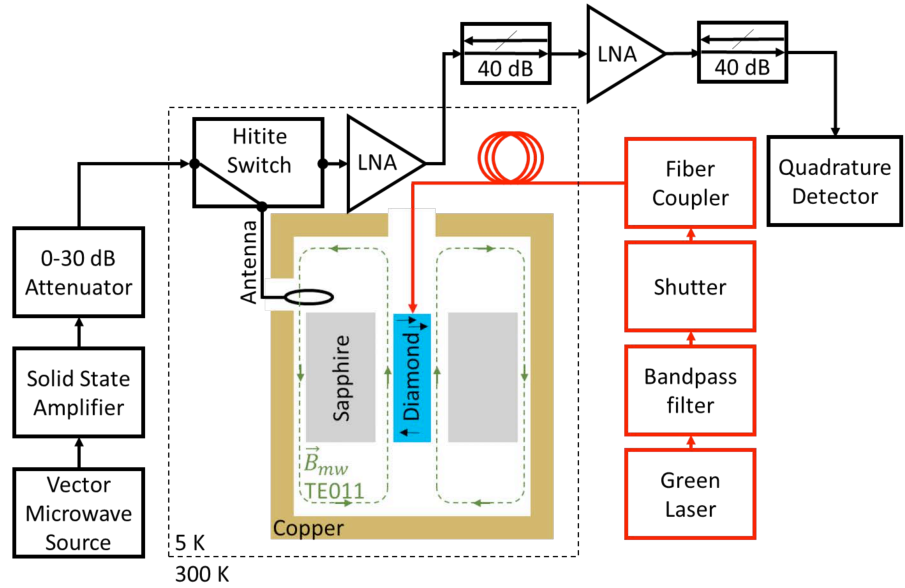

**Fig. S3 Pulsed electron spin resonance experimental set up.**
A vector microwave source is amplified and coupled capacitively to the TE011 mode of a sapphire-dielectric cylindrical resonator. The diamond sample sits in the center of the resonator where the microwave field ($B_{\mathrm{mw}}$) is uniform. The spin echo is coupled out of the resonator through a different path set by a fast Hittite microwave switch and amplified by a low noise cryogenic preamplifier. Optical excitation from a green laser (532 nm) is controlled with a mechanical shutter and fiber coupled to the sample.

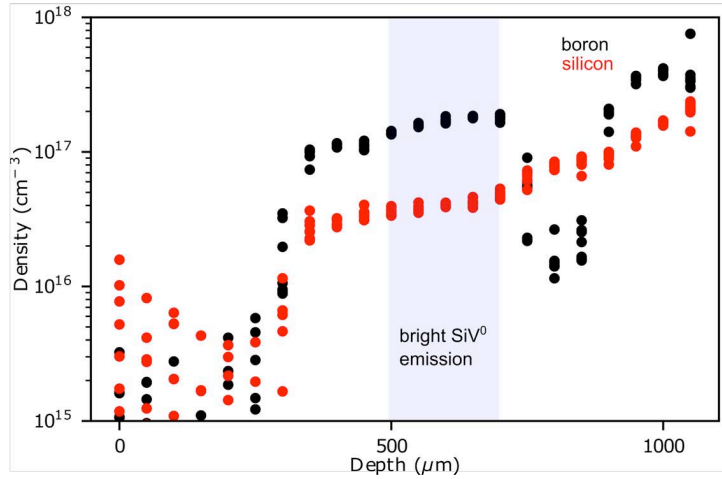

**Fig. S4 Secondary ion mass spectrometry elemental survey of boron and silicon.**
The elemental composition of boron and silicon was determined along the growth axis of the modulation doped diamond (layered sample) using secondary ion mass spectroscopy. The region from 0-300 μm corresponds to the HPHT region and the region from 300-1100 μm is the CVD growth. The concentrations of both species were increased by about one order of magnitude throughout the CVD growth. The shaded region shows where $SiV^0$ PL is observed.

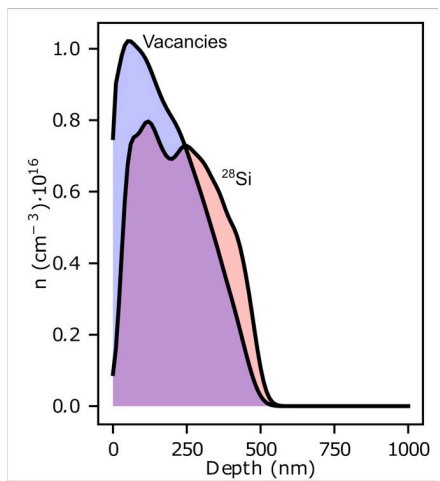

**Fig. S5 Implantation profile for the implanted sample.**
Implantation profiled of $^{28}$Si (red) and damage induced vacancies (blue) for the seven-step implantation recipe in Table S1 calculated with SRIM simulations. The volume of the implanted region was maximized to produce an ideal sample for bulk measurement.

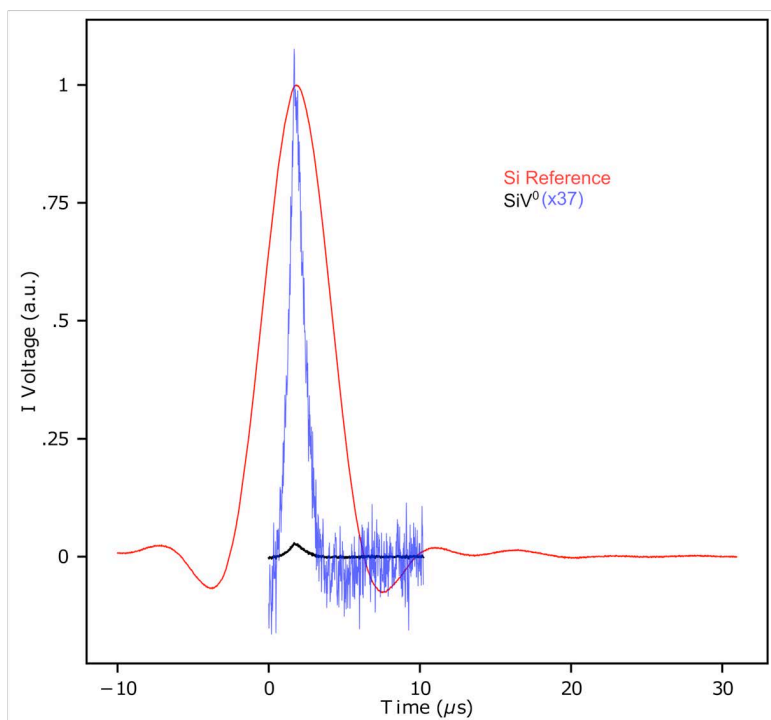

**Fig. S6 Spin counting experiment for determining the number of SiV$^0$ in the implanted sample.**

Spin counting experiment comparing the Hahn echo intensity in the implanted SiV$^0$ sample (black curve) with the echo intensity of a silicon sample (red curve) with a known number of phosphorus donor electron spins measured under the same experimental conditions. The interpulse delay, $\tau$, was set shorter than $T_2$. The blue curve is the black curved multiplied by a factor of 37 to match the amplitude of the reference sample. The comparison gives an estimated conversion efficiency of 90±10%. The baseline of the SiV$^0$ trace is cut off on the left-hand side, corresponding to where the microwave switch is moved to the detecting position. This was necessary due to the large quality factor needed to maximize the sensitivity of the measurement of the implanted sample.

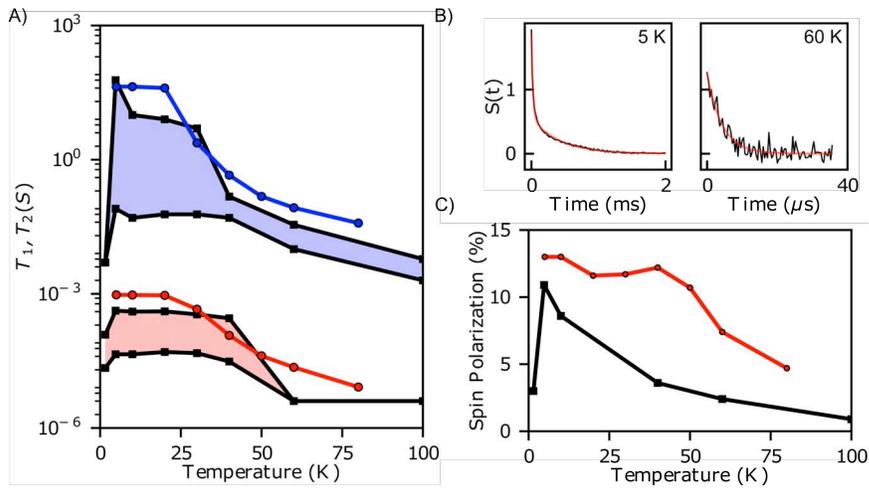

**Fig. S7 Pulsed electron spin resonance measurements of the layered sample**

(A) Temperature dependence of $T_1$ (blue region) and $T_2$ (red region) for layered sample as extracted from a biexponential fit. The shaded regions represent the distribution of decay times present in this sample, $[T_f, T_s]$. For comparison, the $T_1$ (blue dots) and $T_2$ (red dots) for the uniformly implanted sample (Fig. 2) are included. (B) Hahn echo decay curves for the layered sample plotted along with their biexponential fits. (C) The temperature dependence of the optically induced spin polarization at 5 K with 532 nm excitation for the layered sample (black dots) and the uniformly implanted sample (red dots).

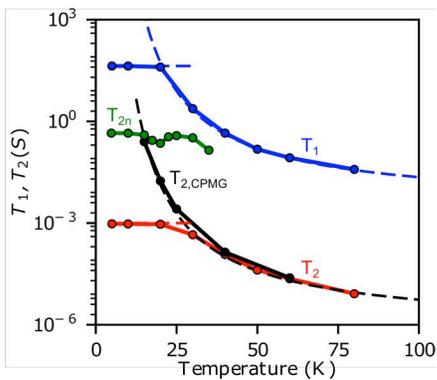

**Fig. S8 Electron nuclear double resonance (ENDOR) measurements of the $^{29}$Si nuclear spin**

ENDOR was used to measure the nuclear spin coherence time (green dots, $T_{2n}$) of the intrinsic $^{29}$Si nuclear spin of SiV$^0$ ($I = \frac{1}{2}$). We observe $T_{2n} = 0.45 \pm 0.03$ s, with a flat temperature dependence at low temperature ($T < 15$ K).

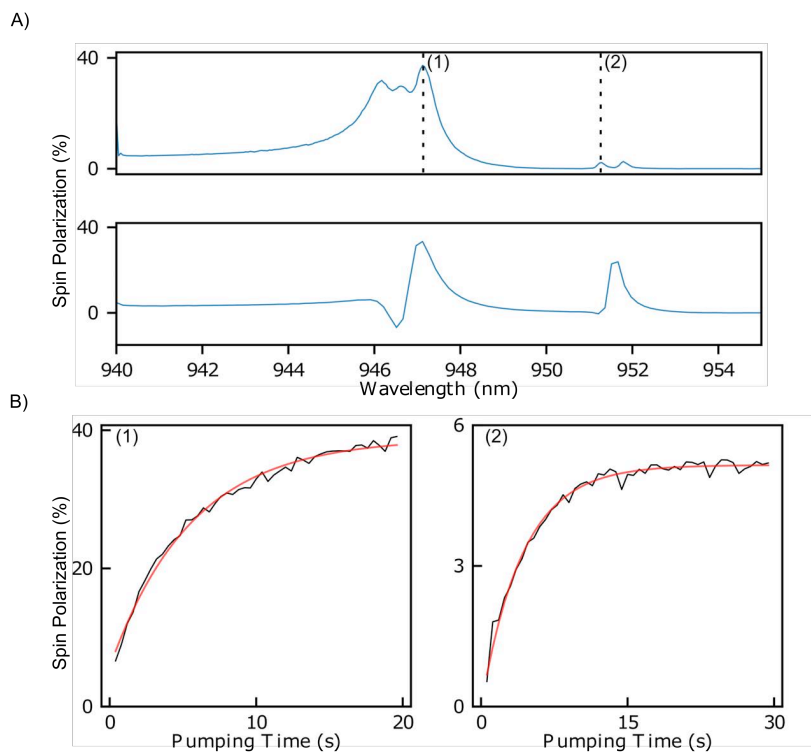

**Fig. S9 Polarization saturation curves with resonant optical excitation.**

(A) Optical spin polarization of the $m_s = 0 \leftrightarrow +1$ transition for a site aligned along the magnetic field direction. The top panel corresponds to a site with a symmetry axis parallel to the magnetic field and the bottom panel corresponds to a site aligned at 109$^o$ to the magnetic field. (B) Accumulation of spin polarization with excitation at 947 nm (curve 1) and 952 nm (curve 2).

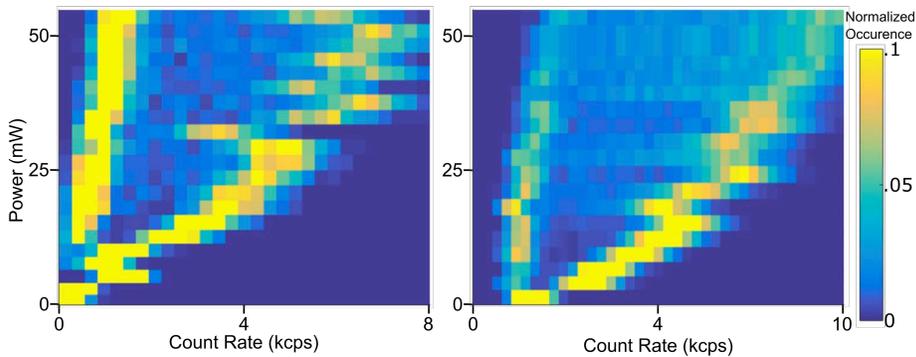

**Fig. S10 Histogram of surface emission rate as a function of laser power**

Normalized distribution in the emission rate of single bright SiV$^0$ as a function of excitation power (905 nm) for two different centers. Both centers exhibit binary switching in the photoluminescence rate as evidenced by the bimodal nature of the distributions. For each slice the peak at the lower count rate corresponds to the off (dark) state of the center combined with a background that linearly increases with laser power. The peak with the higher count rate corresponds to the on (bright) state of the SiV$^0$ center and the count rate shows some saturation with the laser power.

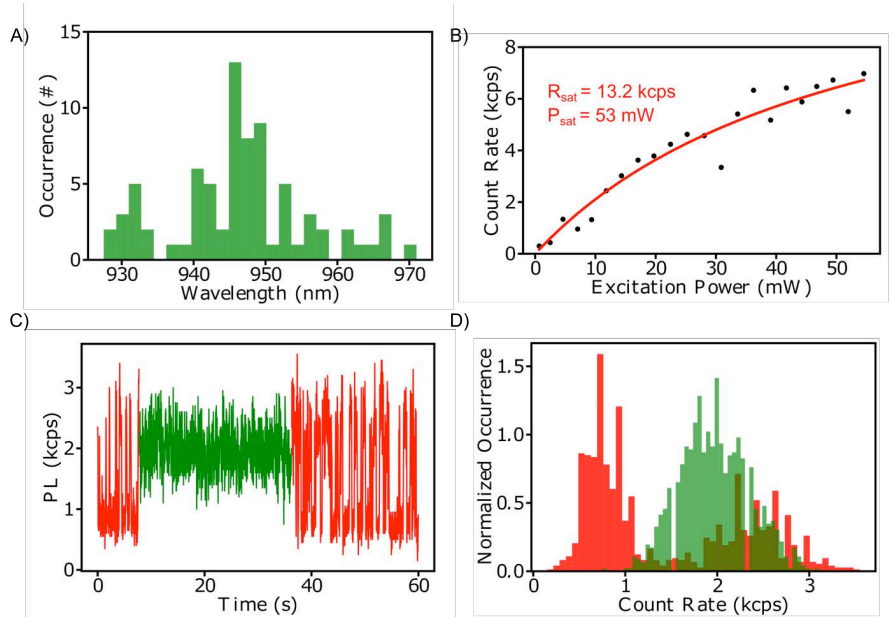

**Fig. S11 Emission characteristics of single centers**

(A) Inhomogeneous distribution of ZPL emission from bright emitters in the implanted sample. (B) Saturation curve of a single center exciting with a 905 nm continuous wave laser. (C) Time trace of photoluminescence count rates from a single center with 905 nm excitation (red points) as well as simultaneous 905 nm and 532 nm excitation (green points). (D) Distribution of count rates from the time trace in C. The green distribution is for the simultaneous illumination and the red distribution corresponds to only 905 nm excitation. The distribution has bimodal character under the presence of only 905 nm excitation, while it appears singly peaked with the addition of 532 nm excitation.

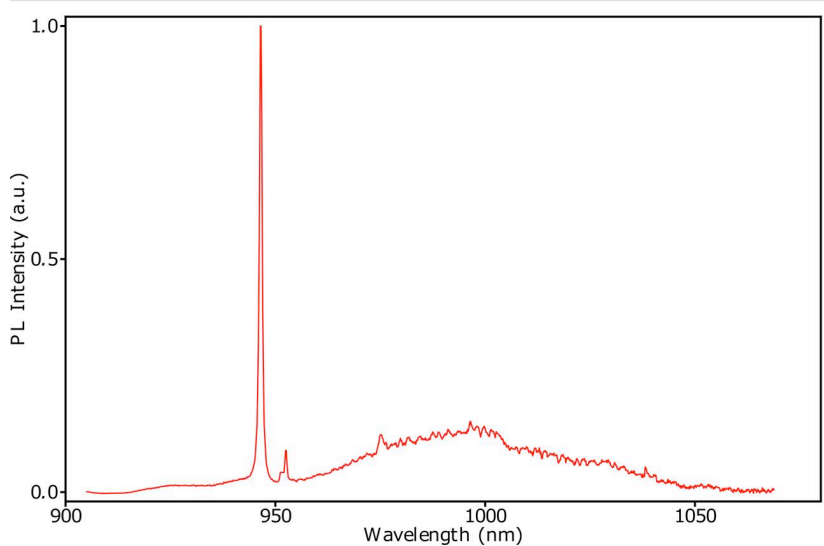

**Fig. S12 Bulk photoluminescence of the $^{29}$Si sample**
Bulk photoluminescence of the $^{29}$Si sample at 4.2 K with 50 mW of excitation at 905 nm. Co-defects near 946 nm obscure measurements of the Debye-Waller factor. The estimated Debye-Waller factor in the $^{29}$Si sample using the PL spectrum is only 0.09.

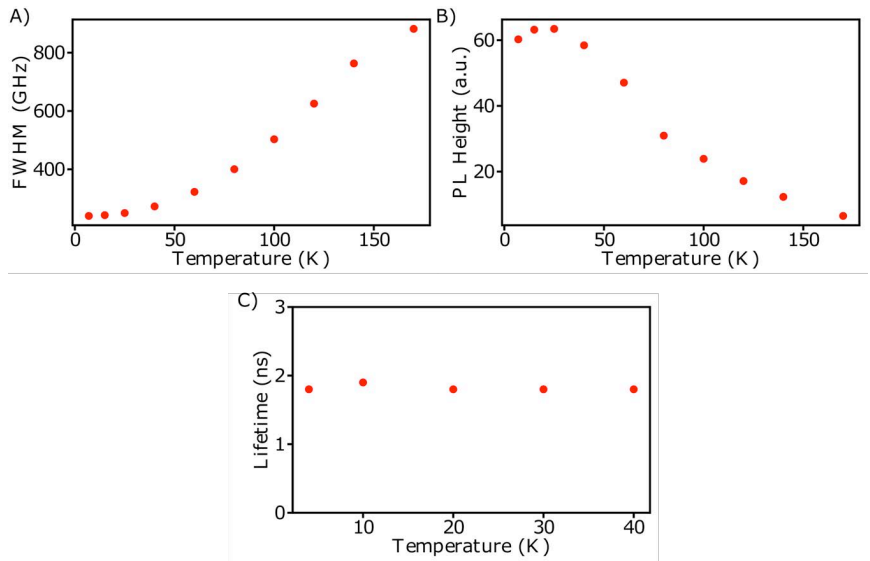

**Fig. S13 Temperature dependence of the bulk SiV⁰ photoluminescence and excited state lifetime**

(A) Ensemble linewidth of the SiV$^0$ zero-phonon line as a function of temperature (4.2-180 K). (B) Photoluminescence intensity (peak height) of the zero-phonon line as a function of temperature. (C) Temperature dependence of the SiV$^0$ excited state lifetime (4.2-40 K).

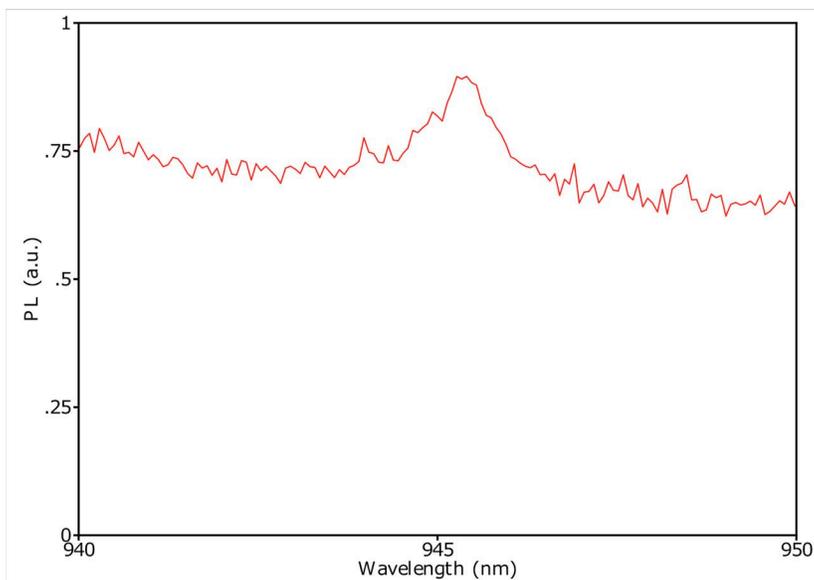

**Fig. S14 Bulk PL of the implanted sample**

Bulk PL of the implanted sample at 77 K with 200 mW of 830 nm excitation. The spectrum exhibits a weak SiV$^0$ ZPL with 1.0 nm linewidth, limited by phonon broadening.

**Table S1 Implantation recipe for the uniformly implanted sample.**

Recipe for implanting a 450 nm, uniform layer of silicon into diamond.

| Step # | Dose ($cm^{-2}$) | Energy (keV) | Tilt (degrees) |
|---|---|---|---|
| 1 | 6.83E10 | 400 | 7 |
| 2 | 5.24E10 | 310 | 7 |
| 3 | 4.67E10 | 240 | 7 |
| 4 | 3.98E10 | 180 | 7 |
| 5 | 3.59E10 | 120 | 7 |
| 6 | 2.85E10 | 80 | 7 |
| 7 | 2.85E10 | 40 | 7 |